\documentclass[10pt, letterpaper, twocolumn, twoside]{IEEEtran}
\usepackage{graphicx,color,subfigure,colordvi}
\usepackage{epsfig}	
\usepackage{textcomp}
\usepackage{verbatim,url}
\usepackage{times}
\usepackage{citesort}

\begin{document}
\title{Model-Based Event Detection in \\ Wireless Sensor Networks}  

\author
{
\centering{
\begin{tabular}[t]{c@{\extracolsep{3em}}c} 
Jayant Gupchup, Randal Burns, Andreas Terzis  &
Alex Szalay \\ 
Department of Computer Science  &
Department of Physics and Astronomy \\
Johns Hopkins University & Johns Hopkins University \\
3400 N. Charles St & 3400 N. Charles St \\
Baltimore, MD 21218 & Baltimore, MD 21218 \\
\texttt{\{gupchup,randal,terzis\}@jhu.edu}  &
\texttt{szalay@jhu.edu} 
\end{tabular}
}
}
\maketitle


%
%

\begin{abstract}
In this paper we present an application of techniques from statistical
signal processing to the problem of event detection in wireless sensor
networks used for environmental monitoring. The proposed approach uses
the well-established Principal Component Analysis (PCA) technique to
build a compact model of the observed phenomena that is able to
capture daily and seasonal trends in the collected measurements. We
then use the divergence between actual measurements and model
predictions to detect the existence of discrete events within the
collected data streams. Our preliminary results show that this event
detection mechanism is sensitive enough to detect the onset of rain
events using the temperature modality of a wireless sensor network.
\end{abstract}

%
%

\section{Introduction}
\label{sec:intro}


A number of testbeds (\emph{e.g.}, \cite{Olin,TPS+05,ucb-hab}) have
shown the potential of wireless sensor networks (WSNs) to collect
environmental data at previously unimaginable spatial and temporal
densities. These developments present many data management challenges.
First, our experience from the deployments has made clear
the shortcomings of the static behavior of current sensor
networks. For example, scientists would like to sample the environment
at a high frequency to capture detailed information about
``interesting'' events, but doing so would create an inordinate amount
of data. On the other hand, sampling at a lower frequency generates
less data but misses important temporal transients. Second, the large
amount of data that these networks generate complicates the querying
and post-processing stages. Rather than manually traversing through
the collected data, scientists would prefer to query for measurements
related with certain events (\emph{e.g.}, significant rainfall). 

To address these issues, we need WSNs that can reason about the
phenomena they observe and change their behavior based on events they
detect.  Possible adaptation strategies include changes in the
sampling rate as well as waking up other nodes in the network to
increase spatial coverage of the detected event~\cite{DGA+05,GS04}.

The readings of sensors are superpositions of several processes.
They are often dominated by predictable foregrounds, which can be very much
larger than the subtle trends and variations that we are trying to measure
or the small events that we try to detect. In order to interpret the readings,
it is important to separate these different signals into independent
components. In environmental monitoring, most sensors witness
daily variations of all quantities
and seasonal trends. In addition, there are
discrete natural events (storm, rainfall, strong winds) that 
have a separable effect on our data. We present an approach
using techniques of statistical signal processing to decompose the
sensor readings into various physically meaningful components.
In our approach, we perform a step-by-step identification of various
foregrounds. We identify the diurnal cycle present in both the box and
soil temperature sensor data and we account for the effect of seasonal
drift. 
We make use of all these priors (daily cycle, seasonal drift) to detect events
by identifying when measurements diverge from those expected by the foregrounds.


Specifically, we explore
variants of Principal Components Analysis (PCA) ~\cite{Duda}
that we use to extract features from the
data collected by the network and discover the multiple underlying
physical processes that generate the observed data. 
This produces a {\em model} of ``normal behavior.''
Observations that diverge from the model correspond well with events.
We note that one can build the PCA model offline using historical data and that a small
number of parameters summarize the phenomena that the motes sense. 
Such a compact representation of the model makes
it possible to build a
lightweight event detection mechanism that runs in real time on the
network's motes.

%

We evaluate the performance of the proposed mechanism using data from
the Life Under Your Feet environmental sensing
network~\cite{Olin}.  We execute the event detection
algorithm to detect rain events with the deployment area 
over ten months of the network's lifetime. 
We compare the list of detected events with precipitation data recorded
by a weather station at BWI airport. 

\begin{figure}
  \centering
  \includegraphics[width=0.9\columnwidth]{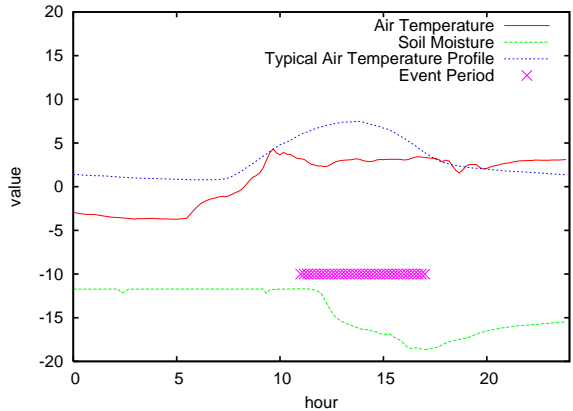}
  \caption{Air temperature is a better indicator of the onset of a
  rain event compared to soil moisture.}
  \label{fig:indicator}
\end{figure}

This specific application reveals
another aspect of the proposed approach: while the motes in our
network have soil moisture sensors, these sensors cannot detect the
onset of a rain event, because soil moisture rises only after the water
seeps through the soil. 
Instead, we use a combination of air and soil
temperature measurements to detect when rain starts to fall. 
Figure \ref{fig:indicator} shows that temperature varies immediately
with the onset of an event, but that soil moisture lags by 
several hours.
The model allows us to detect the rain event rapidly based on indirect
evidence prior to the rain's direct effect on soil moisture.
This better describes system behavior, capturing much more information 
about the dynamics of soil moisture in response to rain.

\subsection{Environmental Sensing}

While our solution is generally applicable to WSNs that collect large
amounts of data using multiple sensing modalities, we present our
design through a environmental monitoring application we developed and
is currently deployed for over 18 months at an urban forest in
Baltimore, MD. The purpose of the \emph{Life Under Your Feet} network
is soil monitoring in which each of the network's ten motes
periodically collects measurements, including soil temperature and
soil humidity, as well as ambient temperature and light.

The key difference between this application and previous environmental
monitoring networks (\emph{e.g.}, \cite{ucb-hab,TPS+05}) is that
\emph{all} raw measurements are reliably retrieved at the network's
base station, which subsequently inserts them to an SQL database. This
stringent reliability requirement is dictated by our scientific
collaborators and the research mission of the monitored site. Each
mote takes measurements at one minute intervals and records them
temporarily to its integrated flash memory. The MicaZ motes we use
have a total capacity of 512 KB of flash storage~\cite{micazspecs}.
In general, each mote stores 23 KB bytes of measurement data per day,
which indicates that measurement data will be lost if not collected
within 20 days. In practice, we download data from each of the
network's motes at least once a week, using an automatic repeat
request (ARQ) protocol to ensure reliable delivery in the presence of
packet losses.

We also extract weather information (air temperature and rain events) 
from a weather station at the BWI airport located 25
miles away from our deployment site.  The data scraping program we
use inserts this data into the same database,
allowing meteorological information, such as rain duration and amount of rainfall,
to be correlated to the
data collected by the sensor network.

%
%

\section{Related work}
\label{sec:related}

PCA event detection constructs a model of system
behavior.  We consider two applications of model-based event
detection in describing related work.
The first is an offline variant in which
event detection happens at the database that stores the
measurements collected by the network and is used to automatically
identify ``interesting'' regions within the swaths of data acquired by
the sensor network. 
The other is online in that motes in the network detect use 
events and models to alter their behavior.

Offline event detection provides a model suitable for querying
events from noisy and imprecise data.  Both database systems
\cite{IBM07,AM06} and sensor networks \cite{AGM+04,JCW04,CHZ02} have
explored model-based queries as a method for dealing with irregular or
unreliable data.  Models in these systems include Gaussian-processes
\cite{AGM+04}, interpolation \cite{GRS00,Neug91}, regression
\cite{AGM+04,GBT+04} and dynamic-probabilistic models
\cite{AM06,JCW04}.  We give another, PCA-based model specifically
suited to event detection.  MauveDB \cite{AM06} provides a user-view
interface to model-based queries, which greatly extends the utility
and usability of models.  We intend to implement our offline PCA model
within the MauveDB framework.

In the online case, sensor networks reduce the bandwidth requirements
of data collection by suppressing results that conform to the model or
compressing the data stream through a model representation.  This has
coincident benefits on resource and energy usage within the network.
If sensors measure spatially correlated values, values collected from
a subset of nodes can be used to materialize the uncollected values
from other nodes \cite{GNDC05,Kotidis05}.  Similarly,
temporally-correlated values may be collected infrequently and missing
values interpolated \cite{JCW04,DYR04}.  By placing models in the mote
itself, the mote may transmit model parameters in lieu of the data,
compressing or suppressing entirely the data stream
\cite{CDHH06,SBF+07,TM06}.  Our PCA model may be used for suppression
and compression and may also be used to alter the behavior and
configuration of the network,
{\em e.g.}~ only collecting data when events occur
and turning off large portions of the network at other times.

Most research on ``event detection'' describes data fusion and
in-network event processing, rather than the detection of an event
based on the data.  REED provides in-network joins to report event
conditions that are programmed declaratively \cite{AML05}.  Other
systems make sure that multiple sensors detect an event prior to
reporting it \cite{LLS+03,HLBJ04}.  Our work focuses on using PCA
models to rapidly and accurately report an event at a single mote.
This single mote report serves as an input to fusion and event query
evaluation.  Other ecological monitoring systems use simple rising
edge or trigger/threshold based event detectors at each mote
\cite{SOP+04}.

We use PCA to determine that a reading or time series is {\em
dissimilar} to the normal behavior of the system, characterize by the
principal components.  Similar uses of PCA
include anomaly and intrusion detection in computer networks
\cite{WGZ04,LakhinaCrovellaDiot:sigcomm05} leakage detection in gas
sensor arrays \cite{PPB+03}.  Recently, PCA has been applied to event
detection in the Internet, specifically identifying correlated
throughput and loss events on multiple Internet paths \cite{slac-pca}.
However, the authors provide no details of their approach.

%
%

\section{Methodology}
\label{sec:sol}



Principal component analysis (PCA) ~\cite{Duda} or Karhunen-Lo\`{e}ve
transform (KLT) is a powerful statistical tool for simplifying data,
by reducing high-dimensional datasets into low-dimensional datasets
that approximate the original data.  It does so through singular value
decomposition (SVD): an orthogonal linear transform of a matrix (the
original data) into an equivalent diagonalized matrix.  The values of
the diagonal matrix are {\em eigenvalues} and the corresponding {\em
eigenvectors} are called basis vectors.  The eigenvectors with maximum
eigenvalues represent the ``most important'' dimensions in that these
dimensions have the maximum variance and strongest correlation in the
dataset.  Thus, the data set may be reduced to just those dimensions
(eigenvectors) with large eigenvalues. 
Data analysis may be performed in the lower dimensional representation 
with good fidelity to results on the original data.  The lower dimensional 
space offers benefits not only in data size, computational complexity, 
and ease of visualization, but also these vectors represent the 
``typical'' patterns of the data, whereas the residuals correspond to 
``atypical'' behavior.  PCA has seen wide-range of applications, including 
clustering, correlation detection, pattern matching, and data compression.

\begin{figure}
\begin{center}
  \includegraphics[width=0.9\columnwidth]{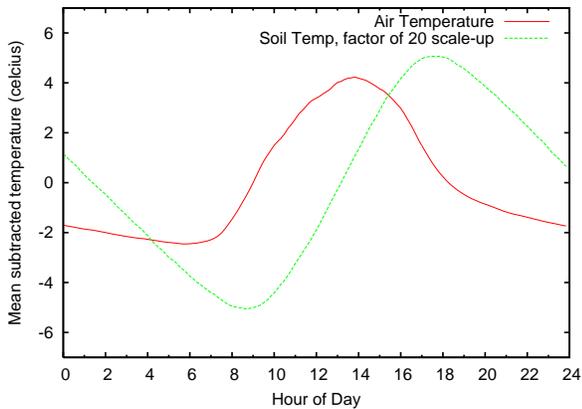}
  \caption {Mean subtracted profile of air and soil temperature (latter 
	scaled up by a factor of 20) for a typical 24 hour cycle} 
  \label{fig:MeanProfile}
\end{center}
\end{figure}


\subsection{Applying PCA to sensor measurements}

We apply our PCA model to air temperature and soil
temperature sensor readings.
Sensor readings exhibit
typical diurnal cycles, which dominate every other signal present. 
Fig \ref{fig:MeanProfile} shows the mean-subtracted profile of a typical day for air 
temperature and soil temperature.  We note the rise in temperature 
as the sun comes out in the morning and the fall in temperature as the 
sun goes down in the evening for air temperature. We also observe that 
soil temperature changes lags air temperature changes by several hours,
owing to the inertia of the soil. There is a noticeable phase shift between 
air temperature and soil temperature. This pattern (AC component) is 
exhibited by all normal (non-event) days of all seasons around the average 
value (DC component) for that day.


LUYF sensors record measurements once every minute. We aggregate 
and smooth multiple readings, which produces a 
data-series with a reading every 10 minutes. 
We find empirically that a 10 minute average reveals useful 
information from the data.  It smooths 
transients, yet samples at a relatively high-frequency. 
This data-series is then
converted into an array of vectors such that each vector represents a
day's readings from midnight to midnight. In a given day, we
have 144, 10 minute intervals. 

\begin{figure}
\begin{center}
  \begin{tabular}{c}
  \epsfig{file=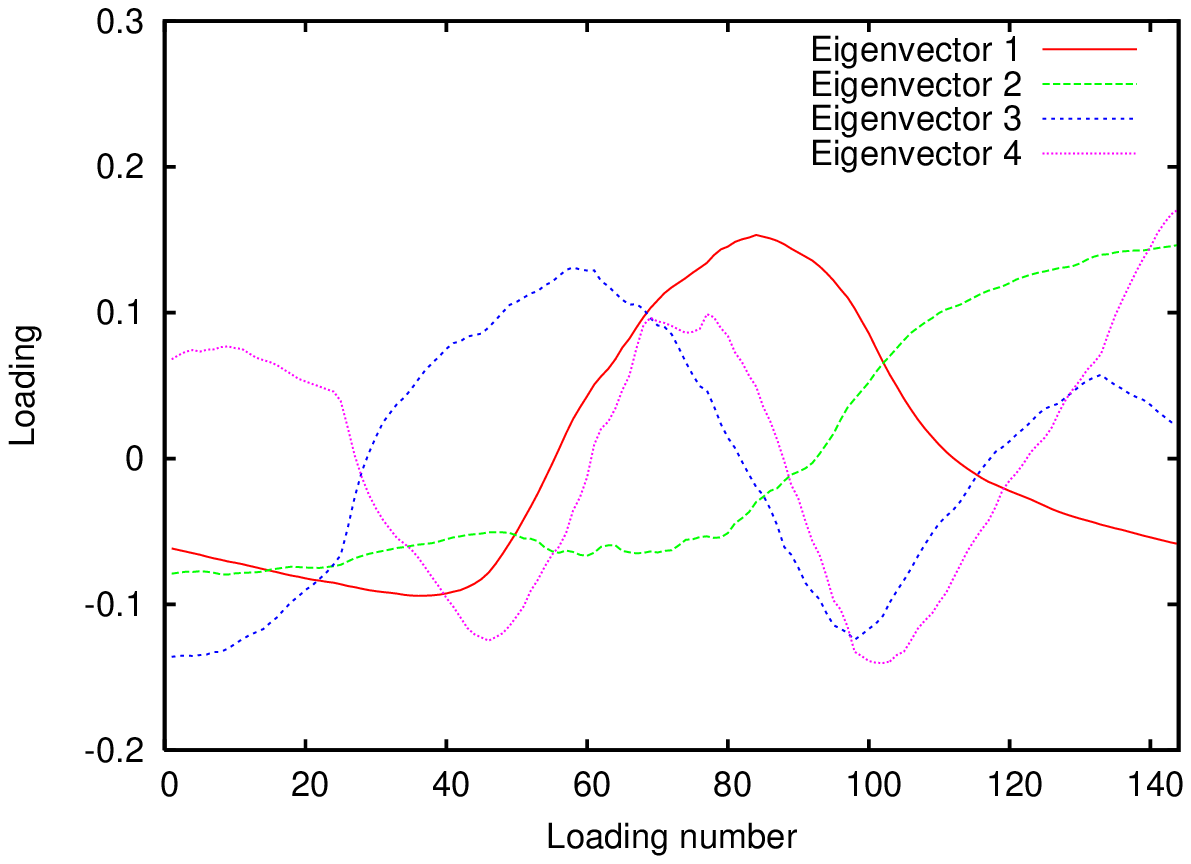, width=0.8\columnwidth} \\
  \epsfig{file=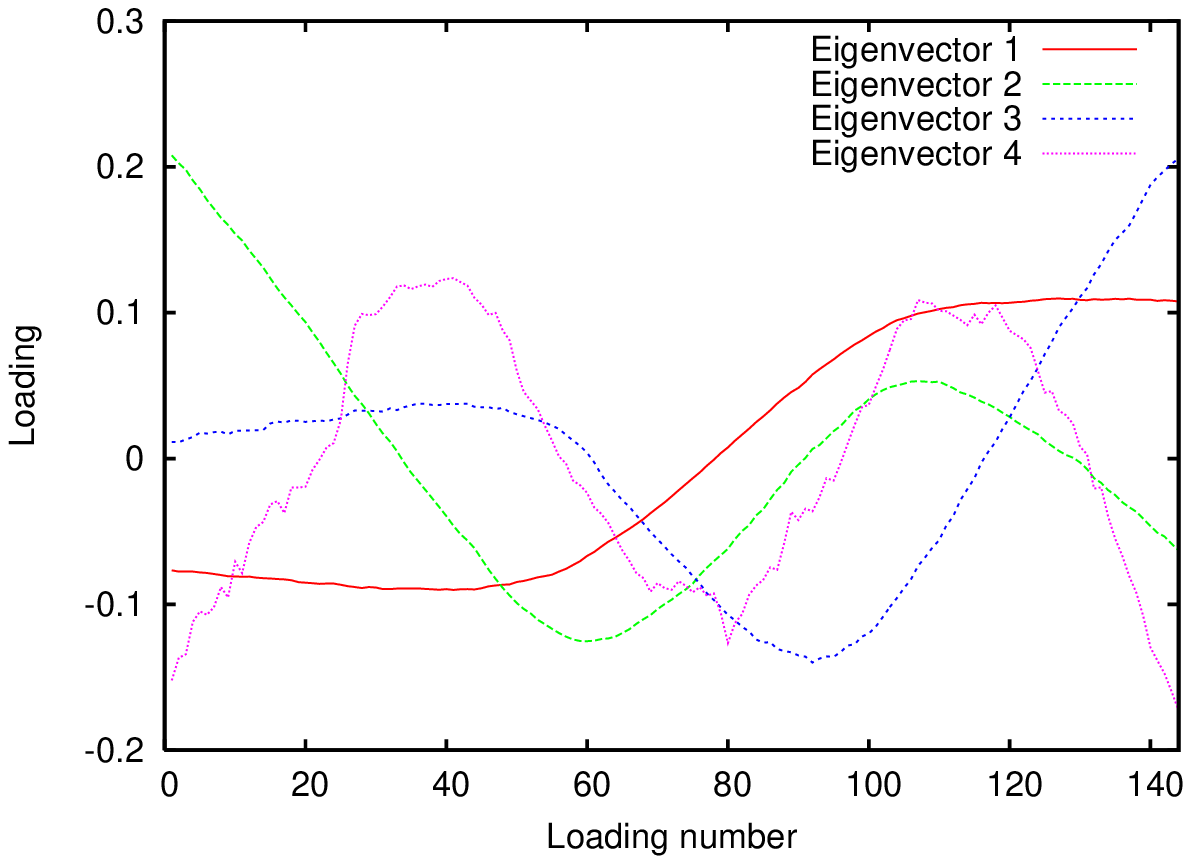, width=0.8\columnwidth} \\
  \end{tabular}
  \caption {Daily temperature eigenvectors in decreasing order of eigenvalues. 
	The top panel shows the basis vectors for the air temperature while the lower 
	panel displays the basis vectors for soil temperature}
  \label{fig:basis}
\end{center}
  \end{figure}

We clean the data prior to building the model in order to best characterize
the ``normal'' behavior of the system.
We subtract the mean temperature
of that given day (calculated separately for each sensor) from each of 
these vectors and normalize the readings in the RMS sense. 
Using normalized vectors ensure that the diagonal elements of the correlation
matrix are unity. Thus, each vector contributes equally to 
the PCA basis.  This balances the contribution of summer and
winter to the model even though summer days have higher variance.
In order to obtain a well-behaved
basis, we censor the days which have a lot of inherent noise and jitter from
our training set. We apply a simple median filter to get rid of these ``bad'' days.

After cleaning the data, we perform a SVD on 
the data to produce our orthogonal eigenvectors (basis vectors) and order 
these vectors by decreasing eigenvalues. Fig ~\ref{fig:basis} shows the basis 
obtained for air temperature and soil temperature for the LUYF deployment between 
the period of September 2005 to July 2006.  We find that the first 4 eigenvectors cover 
90.95\% of the total variation in the air temperature data and 98.89\% in the 
soil temperature data (as defined by the sum of the first four eigenvalues of 
the diagonal matrix divided by the trace). The first eigenvector accounts
for 55\% of the total variation in the air temperature data. 
The physical meaning of the different eigenvectors are apparent. 
The first component of the air temperature is a bell shape curve, 
corresponding to the slow rise of the temperature around 7 am, then cooling 
after 3pm. The second eigenvector is rising throughout the day monotonically, 
describing a warming/cooling trend from one day to another. The third vector
causes the bell shaped curve of the temperature to slide forward or backward,
representing the effect of the seasonal warming and cooling.
Finally, the fourth eigenvector is the broadening and shortening of
the daily temperature cycle, again a seasonal effect. 

The soil has a large inertia in responding to changes in the external temperature,
the characteristic timescale is longer than a day. This manifests itself in the 
fact that the most significant eigenvector is the cooling/warming, and all others
(daily cycle, shift and broadening) are substantially suppressed in amplitude
and have a significant phase shift.

\subsection{Expansion on the Basis and Long-Term Trends}

To complete the model, we factor in the contributions of all sensors
over all time.
We expand all the daily vectors over the basis vectors. This gives us 
four coefficients ($e_{i1},...,e_{i4}$) to describe the daily behavior of the 
temperature for each $sensor_{i}$ (five, if we add the mean temperature as $e_{i0}$).
In order to identify long-term trends, we iteratively run a low-pass filter with a 
fixed width of one week over the different series, resulting in the smooth 
series $s_{i0},.., s_{i4}$.  For each of those coefficients we average
over all sensors to get the smooth mean ($S_{0},.., S_{4}$). Hereafter, we will use 
capitals to denote a time series averaged over all the sensors.

The smoothed series exhibit strong correlations. 
$S_{3}$ and $S_{4}$ describe the beginning and the length of daytime,
whereas $S_{2}$ describes the slow warming and cooling trends, associated with 
the changes of seasons. These smooth trends serve as the background to all 
the other variations. 

\subsection{Event detection}
\label{sec:offline}

Our general approach to event detection looks at the coefficient
of the first eigenvector.
We began by looking at the projections of each day's
mean-subtracted air temperature on the first few eigenvectors. Although the 
first 4 eigenvectors for air temperature represent 90.95\% of the
total variation in the data, we realized that most of the information
is shown by the coefficient of first eigenvector. Thus, we were able 
to analyze an entire day's data by looking at the series $e_{i1}$  thereby 
achieving a massive compression. We created the data series $E_{1}$, 
the eigen-coefficient $e_{1}$ for that day averaged over all sensors. 
We applied a threshold on the $E_{1}$ series to detect events: low values
correspond to behavior that differs from the model. We refer to this 
method as the \textsc{Basic} method. Although this approach gave us satisfactory results, 
it does not take into account the seasonal drift. 

We improve on the \textsc{Basic} detector by removing the seasonal 
drift and running a high pass filter on the $e_{i1}$ data series.
We run the high-pass filter using the 
difference $D_{1} = E_{1}-S_{1}$ between the data series $E_1$ and the smoothed series
$S_1$.
We refer to this method as the \textsc{Highpass} method. 
It significantly improves the number of events detected and reduces the
number of false negatives. 

The last approach we present makes use
of the inertia exhibited by the soil temperature. Since soil
temperature changes much slower compared to the air temperature, we
looked at the differences between the high-pass filtered series, $D_{1}$ for
air temperature and the high-pass filtered data series, $D_{1}$ for soil
temperature and then set a suitable threshold for detecting events. We
refer to this approach as the \textsc{Delta} method. It significantly 
outperforms the \textsc{Basic} and the \textsc{Highpass} methods. 
We find that because of the 
inertia shown by soil temperature, the eigen-coefficients $E_{1}$ for soil temperature 
show sharp changes on the day(s) after the event. This made the event 
days easier to identify.

%
%

\section{Evaluation}
\label{sec:eval}

\begin{figure}
\begin{center}
  \includegraphics[width=0.9\columnwidth]{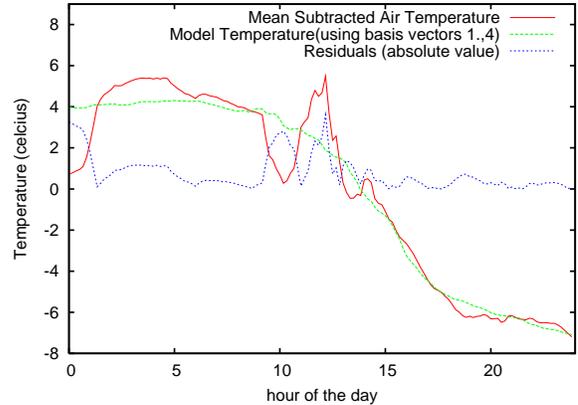}
  \caption {Difference between Air temperature measurement and model projection for the rain event on 2006-01-18}
  \label{fig:event_residual}
\end{center}
\end{figure}

We use our model to detect events on the deployment for the
period between September 2005 and August 2006 and compare the results with the actual known
events recorded by a weather station at Baltimore-Washington
International (BWI) airport ~\cite{BWI}.  We assume that rain at 
BWI implies rain at Johns Hopkins University, Baltimore which is located 
25 miles away. In our evaluation, we only consider rain 
events which are prominent. For example, we consider event days as 
days having precipitation greater than 3 mm. We
considered 225 days starting from September 17, 2005 and July 20, 2006, and found that 48
events fit this criteria. 

There are many other types of events which have
also occurred during the days of our sampling: faulty sensors,
motes running out of power, etc.
Particularly interesting was a period of
about 45 days from mid March 06 to the end of April 06 in which there were
lots of anomalies in the $e_{1}$ values. This was the result of sporadic
direct sunlight heating up the motes. After April, there was enough 
foliage cover that the motes (located at ground level) were not exposed
to the direct heating of the sun.

We focus on the efficiency of detecting the rain events just from
temperature data. There is a good physical basis for this: during rainfall
the temperature suddenly drops, but once the rain is over it recovers.
This produces a large transient on the shape of the 24 hour cycle for that 
particular day, resulting in a smaller $e_{1}$ coefficient and a larger 
residual. Fig ~\ref{fig:event_residual} illustrates this fact. We observed a major
event on 2006-01-18. There was heavy rain between 9:00 AM and 11:00 AM. We can 
clearly see large residuals for this period.  

\begin{figure}
\begin{center}
  \includegraphics[width=0.9\columnwidth]{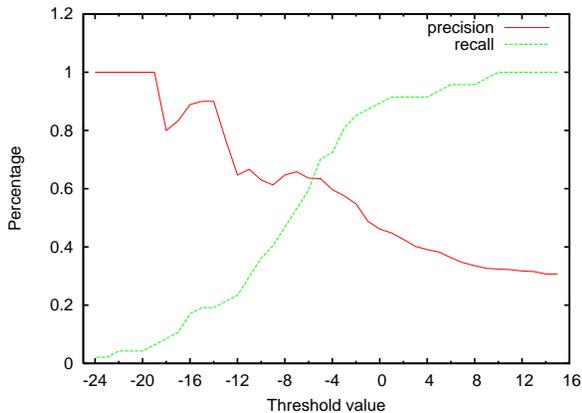}
  \caption {Precision-recall curve for the Delta method}
  \label{fig:prec-recall}
\end{center}
\end{figure}

\begin{table}
\caption{Performance of different methods for detecting events.} 
\begin{center}
\begin{tabular}{|c|c|c|c|}
\hline
Method  &  Precision  &  Recall  &  False Negatives \\
\hline
\textsc{Basic} & 52.459\% & 64\% & 18 \\  
\hline 
\textsc{Highpass} & 51.28\% & 80\% & 10 \\
\hline
\textsc{Delta} & 54.795\% & 85.106\% & 7 \\
\hline
\end{tabular}
\label{tab:perf}
\end{center}
\end{table}

We evaluate the performance of the three methods i.e. \textsc{Basic}, \textsc{Highpass}
and \textsc{Delta} method. In our evaluation, we use the standard information 
retrieval measures of precision and recall.  In this case, precision is the fraction 
of reported events that were actually rain events and recall is the fraction of rain
events that the PCA model reported correctly.  We also report false negatives,
which effect recall and not precision.
We attempt to strike a balance between precision and
recall. Our criteria is to detect as many events as possible with a
true positive rate (precision) of at least 50\%.   Higher precision is difficult
to achieve given that our system also detects other (non-rain) events.
Recall may be affected by the assumption that rain at BWI implies rain at JHU and vice
versa.  This is not always the case.
Table ~\ref{tab:perf} shows the results for the different methods.
Using high-pass filtering and including soil temperature increases recall
without affecting precision substantially.

\begin{figure}
\begin{center}
  \includegraphics[width=0.9\columnwidth]{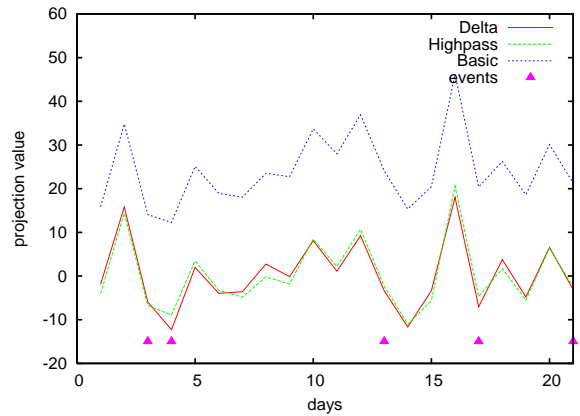}
  \caption {Projection values for different techniques on event and
  non-event days. The marker at the bottom indicates an event}
  \label{fig:projection}
\end{center}
\end{figure}

Figure ~\ref{fig:projection} shows the projection
values of different methods for the period between 12/13/2005 and
01/02/2006. The rain events are indicated by a triangular marker at the
bottom. We can see that the \textsc{Delta} method shows sharper negative peaks
than the other methods on event days and shows lower peaks for
non-event days.  
Notice that the large downward spike shown on day 4 (12/16/2005) 
corresponds to a large event.

We are able to detect most events days, missing only 7 with the
\textsc{Delta} method.  Again, we focus on recall, given that
non-rain events occur and pollute our precision statistics.
The precision-recall curves for different threshold values 
(Figure \ref{fig:prec-recall}) shows that good recall can be achieved
at better than 50\% precision.  The converse is not true.
High recall matches well with our application needs; reporting 
events when they occur supports network adaptation and 
identifies interesting regions of data to scientists.
In all likelihood, precision and recall would be much improved 
with more accurate and local weather monitoring -- a better
``ground truth'' -- and considering multiple types of events.


%
%

\section{Discussion and Future Work}
\label{sec:sum}


In this paper we present an application of techniques from statistical
signal processing to detect the presence of events (\emph{e.g.}, rain
events) that deviate from the regular physical patterns witnessed by a
sensor network. We do this by using a variant of the Principal
Component Analysis (PCA) technique to generate a compact profile for
'normal' measurements. We can then compare actual mote measurements
with model predictions and classify the instances in which the two
diverge significantly as events of interest. We evaluate the
performance of the proposed mechanisms using temperature measurements,
collected over a year by a small environmental monitoring network, to
detect the onset of rain events. Our preliminary results show that
this technique is able to detect most rain events, with small number
of false positives, even in the presence of large foreground
variations and a substantial seasonal drifts.


This is only the beginning---one can carry this approach much
further. While we present event detection in its offline setting, the
observation that only a small number of components can accurately
describe the collected data suggests that the same mechanism can be
implemented on the network's motes. This in turn can result in a
light-weight adaptive sampling algorithm that will enable real-life
WSN deployments confronted with slowly varying environments as well as
sudden, discrete events. Efficient event detection is at the core of
any adaptive observing strategy, and we demonstrate how this can be
done on today's WSN platforms.


At this point the method is able to detect global events, \emph{i.e.}
events that all the sensors experience. However, one would like to
detect localized events. While it is seemingly possible to apply the
same PCA technique to detect events experienced by a single mote, it
becomes harder to differentiate between an actual event and a
malfunctioning sensor.  The question is then how much additional
information is necessary to separate faults from actual events. The
sensors are expected to have variations due to their local environment
(located near/far from a stream, sitting on a hillside with a steep
gradient, etc.) which will cause small, but consistent, correlated
changes. The task is then to find groups of sensors with correlated
measurements. We can do so by removing the obvious daily foregrounds,
and the long seasonal trends, at which point we expect to see these
small correlated differences in the behavior of sensors in the same
group. Once such groups are created, we can compare the projected
measurements of a mote with the measurements of other group
members. If those agree, then a localized event is most likely
occurring, otherwise one (or more) of the sensors are faulty.



So far, we completely exclude from the training set, days with partial
data in which due to some hardware errors we did not get a reading for
every one of the 144 sampling periods. However, it is easy to apply a
''gappy'' Karhunen-Loeve transformation \cite{gappy}, in which the
expansion coefficients can still be computed over a partial
support. Doing so, will enable the creation of a more representative
compressed model of the measurement data and hopefully lead to higher
detection accuracy.


%
%

\section*{Acknowledgments}
\label{sec:ack}

We would like to thank Ching-Wa Yip (JHU, Department of Physics and
Astronomy) for making available to us her PCA C\# library and
providing us her valuable time in the discussions. The data collected
here was done in collaboration with Katalin Szlavecz (JHU, Department
of Earth and Planetary Science) and Razvan Musaloiu-E (JHU, Department
of Computer Science) . The on-line database was built in collaboration
with Jim Gray and Stuart Ozer (Microsoft Research). Their help and
contributions are gratefully acknowledged.

\bibliographystyle{IEEEtran}
\bibliography{sensors}

\begin{thebibliography}{10}
\providecommand{\url}[1]{#1}
\csname url@rmstyle\endcsname
\providecommand{\newblock}{\relax}
\providecommand{\bibinfo}[2]{#2}
\providecommand\BIBentrySTDinterwordspacing{\spaceskip=0pt\relax}
\providecommand\BIBentryALTinterwordstretchfactor{4}
\providecommand\BIBentryALTinterwordspacing{\spaceskip=\fontdimen2\font plus
\BIBentryALTinterwordstretchfactor\fontdimen3\font minus
  \fontdimen4\font\relax}
\providecommand\BIBforeignlanguage[2]{{%
\expandafter\ifx\csname l@#1\endcsname\relax
\typeout{** WARNING: IEEEtran.bst: No hyphenation pattern has been}%
\typeout{** loaded for the language `#1'. Using the pattern for}%
\typeout{** the default language instead.}%
\else
\language=\csname l@#1\endcsname
\fi
#2}}

\bibitem{Olin}
R.~{Mus\u aloiu-E.}, A.~Terzis, K.~Szlavecz, A.~Szalay, J.~Cogan, and J.~Gray,
  ``{Life Under Your Feet: A Wireless Soil Ecology Sensor Network},'' in
  \emph{Proceedings of the Third Workshop on Embedded Networked Sensors (EmNets
  2006)}, May 2006.

\bibitem{TPS+05}
G.~Tolle, J.~Polastre, R.~Szewczyk, N.~Turner, K.~Tu, P.~Buonadonna,
  S.~Burgess, D.~Gay, W.~Hong, T.~Dawson, and D.~Culler, ``{A Macroscope in the
  Redwoods},'' in \emph{Proceedings of the Third ACM Conference on Embedded
  Networked Sensor Systems (SenSys)}, Nov. 2005.

\bibitem{ucb-hab}
A.~Mainwaring, J.~Polastre, R.~Szewczyk, D.~Culler, and J.~Anderson, ``Wireless
  sensor networks for habitat monitoring,'' in \emph{Proceedings of 2002 ACM
  International Workshop on Wireless Sensor Networks and Applications}, Sept.
  2002.

\bibitem{DGA+05}
P.~Dutta, M.~Grimmer, A.~Arora, S.~Bibyk, and D.~Culler, ``Design of a wireless
  sensor network platform for detecting rare, random, and ephemeral events,''
  in \emph{Proceedings of IPSN}, 2005.

\bibitem{GS04}
L.~Gu and J.~Stankovic, ``Radio triggered wake-up capability for sensor
  networks,'' in \emph{Real-Time Applications Symposium}, 2004.

\bibitem{Duda}
R.~Duda, P.~Hart, and D.~Stork, \emph{Pattern Classification}.\hskip 1em plus
  0.5em minus 0.4em\relax Wiley, 2001.

\bibitem{micazspecs}
C.~Corporation, ``{MICAz Specifications},'' Available at
  \url{http://www.xbow.com/Support/Support_pdf_files/MPR-MIB_Series_Users_Manu%
al.pdf}.

\bibitem{IBM07}
IBM, ``Db2 intelligent miner,'' 2007, available at
  \url{http://www-306.ibm.com/software/data/iminer/}.

\bibitem{AM06}
A.~Deshpande and S.~Madden, ``Mauvedb: supporting model-based user views in
  database systems,'' in \emph{Proceedings of ACM SIGMOD}, 2006.

\bibitem{AGM+04}
A.~Deshpande, C.~Guestrin, S.~Madden, J.~M. Hellerstein, and W.~Hong,
  ``Model-driven data acquisition in sensor networks,'' in \emph{Proceedings of
  VLDB}, 2004.

\bibitem{JCW04}
A.~Jain, E.~Change, and Y.~Wang, ``Adaptive stream resource management using
  kalman filters,'' in \emph{Proceedings of ACM SIGMOD}, 2004.

\bibitem{CHZ02}
M.~Chu, H.~Haussecker, and F.~Zhao, ``{Scalable information-driven sensor
  querying and routing for ad hoc heterogeneous sensor networks},''
  \emph{International Journal of High-Performance Computing Applications},
  vol.~16, no.~3, 2002.

\bibitem{GRS00}
S.~Grumbach, P.~Rigaux, and L.~Segoufin, ``Manipulating interpolated data is
  easier than you thought,'' in \emph{Proceedings of VLDB}, 2000.

\bibitem{Neug91}
L.~Neugebauer, ``Optimization and evaluation of database queries including
  embedded interpolation procedures,'' in \emph{Proceedings of SIGMOD}, 1991.

\bibitem{GBT+04}
C.~Guestrin, P.~Bodik, R.~Thibaux, M.~Paskin, and S.~Madden, ``{Distributed
  Regression: an Efficient Framework for Modeling Sensor Network Data},'' in
  \emph{Proceedings of IPSN 2004}, Apr. 2004.

\bibitem{GNDC05}
H.~Gupta, V.~Nacda, S.~Das, and V.~Chowdhary, ``Energy-efficient gathering of
  correlated data in sensor networks,'' in \emph{Proceedings of MobiHoc}, 2005.

\bibitem{Kotidis05}
Y.~Kotidis, ``Snapshot queries: towards data-centris sensor networks,'' in
  \emph{Proceedings of ICDE}, 2005.

\bibitem{DYR04}
A.~Deligiannakis, Y.~Kotidis, and N.~Roussopoulos, ``Compressing historical
  information in sensor networks,'' in \emph{Proceedings of SIGMOD}, 2004.

\bibitem{CDHH06}
D.~Chu, A.~Deshpande, J.~Hellerstein, and W.~Hong, ``Approximate data
  collection in sensor networks using probabilistic models,'' in
  \emph{Proceedings of ICDE}, 2006.

\bibitem{SBF+07}
A.~Silberstein, R.~Braynard, G.~Filpus, G.~Puggioni, A.~Gelfand, K.~Munagala,
  and J.~Yang, ``Data-driven processing in sensor networks,'' in
  \emph{Proceedings of Conference on Innovative Data Systems Research}, 2007.

\bibitem{TM06}
D.~Tulone and S.~Madden, ``{PAQ}: Time series forecasting for aproximate query
  answering in sensor networks,'' in \emph{Proceedings of the European
  Conference on Wireless Sensor Networks}, 2006.

\bibitem{AML05}
D.~J. Abadi, S.~Madden, and W.~Lindner, ``Reed: Robust, efficient filtering and
  event detection in sensor networks,'' in \emph{Proceedings of VLDB}, 2005.

\bibitem{LLS+03}
S.~Li, Y.~L. adn S.~H.~Son, J.~A. Stankovic, and Y.~Wei, ``Event detection
  services using data service middleware in distributed sensor networks,'' in
  \emph{Proceedings of IPSN}, 2003.

\bibitem{HLBJ04}
A.~Herbold1, T.~Lamarre, N.~Bulusu, and S.~Jha, ``Resilient event detection in
  wireless sensor networks,'' in \emph{Proceedings of Intelligent Sensors,
  Sensor Networks and Information Processing}, 2004.

\bibitem{SOP+04}
R.~Szewczyk, E.~Osterweil, J.~Polastre, M.~Hamilton, A.~Mainwaring, and
  D.~Estrin, ``Habitat monitoring with sensor networks,'' \emph{CACM}, vol.~47,
  no.~6, 2004.

\bibitem{WGZ04}
W.~Wang, X.~Guan, and X.~Zhang, ``A novel intrusion detection method based on
  principle component analysis in computer security,'' in \emph{Proceedings of
  Advanced in Neural Networks}, 2004.

\bibitem{LakhinaCrovellaDiot:sigcomm05}
\BIBentryALTinterwordspacing
A.~Lakhina, M.~Crovella, and C.~Diot, ``Mining anomalies using traffic feature
  distributions,'' in \emph{Proceedings of ACM SIGCOMM 2005}, Aug. 2005, pp.
  217--228. [Online]. Available:
  \url{http://www.cs.bu.edu/faculty/crovella/paper-archive/sigc05-mining-anoma%
lies.pdf}
\BIBentrySTDinterwordspacing

\bibitem{PPB+03}
A.~Perera, N.~Papamichail, N.~B\^arsan, U.~Weimar, and S.~Marco, ``On-line
  event detection by recursive dynamic principal component analysis and gas
  sensor arrays under drift conditions,'' in \emph{Proceedings of IEEE
  Sensors}, 2003.

\bibitem{slac-pca}
A.~Iqbal, ``Multi-route event detection using {PCA},'' 2006, available at --
  \url{http://confluence.slac.stanford.edu/display/IEPM/Multi-route+Event+Dete%
ction+Using+PCA}.

\bibitem{BWI}
``{Baltimore-Washington International airport, weather station},'' Available
  at:
  \url{http://weather.marylandweather.com/cgi-bin/findweather/getForecast?quer%
y=BWI}.

\bibitem{gappy}
A.~J. {Connolly} and A.~S. {Szalay}, ``{A Robust Classification of Galaxy
  Spectra: Dealing with Noisy and Incomplete Data},'' \emph{The Astronomical
  Journal}, vol. 117, pp. 2052--2062, May 1999.

\end{thebibliography}

\end{document}